\documentclass[aps,prl,twocolumn,groupedaddress]{revtex4-1}
\usepackage{graphicx}  
\usepackage{bm}        
\usepackage{amssymb}   
\usepackage{amsmath}
\usepackage{setspace} 

\usepackage{hyperref}
\hypersetup{
	colorlinks   = true,    
	urlcolor     = blue,    
	linkcolor    = blue,    
	citecolor    = red      
}

\begin{document}

\title{Third-order nonlinearity OPO: Schmidt Mode Decomposition And Tripartite Entanglement}

\author{C. Gonz\'alez-Arciniegas$^{1}$}
\author{Nicolas Treps$^{2}$}
\author{P. Nussenzveig}
\affiliation{$^{1}$Instituto de F\'{\i}sica, Universidade de S\~ao Paulo, 05315-970 S\~ao Paulo, SP-Brazil}
\affiliation{$^{2}$Laboratoire Kastler Brossel, UPMC-Sorbonne Universités, ENS-PSL Research University, Collège de France, CNRS; 4 place Jussieu, F-75252 Paris, France}

\begin{abstract}

We investigate quantum properties of light in optical parametric oscillators (OPOs) based on four-wave mixing gain in media with third-order non-linearities. In spite of other competing $\chi^{(3)}$ effects, such as phase modulation, bipartite and tripartite entanglement are predicted above threshold. These findings are relevant for recent implementations of CMOS-compatible on-chip OPOs.
\end{abstract}

\maketitle

Optical Parametric Oscillators (OPOs) are ubiquitous sources of non-classical light. Typically, they are based on parametric down conversion (PDC) inside a cavity, a second-order non-linear process in which pairs of (signal and idler) photons are created upon annihilation of incident pump photons. Owing to energy conservation and phase matching, signal and idler fields have strong intensity correlations and phase anti-correlations, resulting in EPR-like entangled states~\cite{Reid1988}. Entanglement is a coveted resource in quantum information science and the OPO has been a driver for many advances, such as the deterministic teleportation of coherent states~\cite{Furusawa1998}, the generation of multicolor entanglement~\cite{Coelho2009,Barbosa2010} with potential use for quantum networks~\cite{Kimble2008}, and the generation of continuous-variable (CV) cluster states in the time domain~\cite{Yokoyama2013} as well as in the frequency domain~\cite{Roslund2013}. The CV entangled states are deterministically generated at a rate which is limited by the OPO cavity bandwidth. 

Recent progress in silicon photonics technologies opened up a new avenue for on-chip CV quantum information. The ultra-small footprint of these devices entails large cavity bandwidths and thus high repetition rates for quantum communications applications. Indeed, silicon based micro-cavities have high quality factors ($Q\gtrsim10^6$), are CMOS-compatible, and are also easily connected to commercial fibers. However, second-order non-linearities are precluded by the material's symmetry properties and the strongest non-linearities are of the third order. Four-wave mixing (FWM) can replace PDC as a source of pairs of signal and idler photons, upon annihilation of pairs of pump photons. Several experiments on quantum states of light have been performed using FWM, such as the first experimental demonstration of squeezed states~\cite{Slusher1985}, and more recently the generation of CV entangled states \cite{Boyer2008} among others. Quantum properties of these on-chip devices have started to be tested and the first experimental generation of squeezed states was recently reported~\cite{Dutt2015}. The generation of frequency combs in this platform~\cite{Levy2010} also opens the possibility to have a scalable generator of cluster states for applications to CV one way quantum information~\cite{Menicucci2006}.

In this Letter, we present a theoretical analysis of $\chi^{(3)}$ OPOs, operating above-threshold, with parameters chosen according to current experiments. When limiting operation to the excitation of only three cavity modes, we predict bipartite and tripartite entanglement, even when including intensity dependent phase modulation effects which are of the same order of the relevant FWM process.

The three cavity modes, named pump, signal and idler, have resonant frequencies $\omega_p$, $\omega_s$, and $\omega_i$, respectively. The field in each mode is described by a corresponding annihilation operator $\hat{a}$. The Hamiltonian, in the rotating wave approximation, is given by
\begin{eqnarray*}
	\hat{H}=\hat{H}_0+\hat{H}_{int} \, ,
\end{eqnarray*}
where the free Hamiltonian $\hat{H}_0$ is given by
\begin{eqnarray*}
	\hat{H}_0=\hbar\sum_{j=\{p,s,i\}}\omega_j\hat{a}_j^\dagger\hat{a}_j.
\end{eqnarray*}
Owing to the dispersion in the cavity, resulting from the geometry and the material dispersion, the mode resonances are not, in general, equally spaced. We define the parameter $D_3=2\omega_p-\omega_s-\omega_i$ to quantify this dispersion. This quantity can be easily engineered in the context of silicon (or silicon-nitride) based ring micro cavities by modifying the radius and/or the transversal shape of the ring.

The nonlinear interaction Hamiltonian $\hat{H}_{int}$ is given by
\begin{gather*}
\hat{H}_{int}=-\hbar \eta\left[ \frac{1}{2}\left(\hat{a}^{\dagger}_p\hat{a}^{\dagger}_p\hat{a}_p\hat{a}_p+\hat{a}^{\dagger}_s\hat{a}^{\dagger}_s\hat{a}_s\hat{a}_s+\hat{a}^{\dagger}_i\hat{a}^{\dagger}_i\hat{a}_i\hat{a}_i\right) \right.\nonumber\\ +2\left(\hat{a}^{\dagger}_p\hat{a}^{\dagger}_s\hat{a}_p\hat{a}_s+\hat{a}^{\dagger}_p\hat{a}^{\dagger}_i\hat{a}_p\hat{a}_i+\hat{a}^{\dagger}_s\hat{a}^{\dagger}_i\hat{a}_s\hat{a}_i\right)
+\left. \left(\hat{a}^{\dagger}_s\hat{a}^{\dagger}_i\hat{a}_p\hat{a}_p +\hat{a}^{\dagger}_p\hat{a}^{\dagger}_p\hat{a}_s\hat{a}_i  \right)\right].
\end{gather*} Here $\eta$ is proportional to the $\chi^{(3)}$ coefficient. The first and second terms in $\hat{H}_{int}$ are respectively the self and cross phase modulation (SPM and XPM) terms and they are responsible for intensity dependent shifts of the resonant frequencies. The last term is the FWM term which is responsible for the energy exchange between the three modes. The SPM and XPM terms play importants roles in the oscillation process and also in the noise and entanglement properties of the system. To our best knowledge, there existed no full analysis of quantum entanglement in above-threshold $\chi^{(3)}$ parametric oscillation in the literature which has taken into account all these properties.

This Hamiltonian, together with the non unitary evolution related with losses and the incoming coupled fields into the cavity (vacuum states for the signal and idler fields and a coherent state of frequency $\Omega_p$ for the pump), leads to the following Heisenberg-Langevin equations of motion for the annihilation operators 
{\small \begin{align}
	\frac{d \hat{a}_p}{dt}=&-\left(\Gamma_p+i \Delta_p\right)\hat{a}_p+i\eta\left[\left(\hat{a}_p^{\dagger}\hat{a}_p+2\hat{a}_s^{\dagger}\hat{a}_s+2\hat{a}_i^{\dagger}\hat{a}_i\right)\hat{a}_p+2\hat{a}_p^{\dagger}\hat{a}_s\hat{a}_i\right]\nonumber\\&+\sqrt{2\gamma_p}\hat{a}_p^{in}+\sqrt{2\mu_p}\hat{a}_p^{loss} \nonumber\\ 
	\frac{d \hat{a}_s}{dt}=&-\left(\Gamma_s+i \Delta_s\right)\hat{a}_s+i\eta\left[\left(2\hat{a}_p^{\dagger}\hat{a}_p+\hat{a}_s^{\dagger}\hat{a}_s+2\hat{a}_i^{\dagger}\hat{a}_i\right)\hat{a}_s+\hat{a}_p^{2}\hat{a}_i^{\dagger}\right]\nonumber\\&+\sqrt{2\gamma_s}\hat{a}_s^{in} +\sqrt{2\mu_s}\hat{a}_s^{loss}\nonumber\\ 
	\frac{d \hat{a}_i}{dt}=&-\left(\Gamma_i+i \Delta_i\right)\hat{a}_i+i\eta\left[\left(2\hat{a}_p^{\dagger}\hat{a}_p+2\hat{a}_s^{\dagger}\hat{a}_s+\hat{a}_i^{\dagger}\hat{a}_i\right)\hat{a}_i+\hat{a}_p^{2}\hat{a}_s^\dagger\right]\nonumber\\&+\sqrt{2\gamma_i}\hat{a}_i^{in} +\sqrt{2\mu_i}\hat{a}_i^{loss} \label{eqn:chi3EqnMovModes}
	\end{align}
} where we have replaced operators $\hat{a}_j$ for corresponding slowly varying operators $\hat{a}_j\rightarrow\hat{a}_j e^{-i \Omega_j t}; \; j={p,s,i}$. The $\Omega_j$ are the frequencies of the carriers which fulfill energy conservation $2\Omega_p-\Omega_s-\Omega_i=0$. The terms $\Delta_j\equiv\omega_j-\Omega_j$ are the field detunings relative to the cavity resonances. The quantities $\gamma_j$ and $\mu_j$ represent the rates for the cavity coupling to output modes and for other losses (scattering, absorption etc.). Total losses are given by $\Gamma_j=\gamma_j+\mu_j$. As long as the cavity modes have similar field profiles, we can set $\Gamma_p=\Gamma_s=\Gamma_i=\Gamma$ (and likewise for $\gamma_j$ and $\mu_j$).  The operators $\hat{a}^{in}$ and $\hat{a}^{loss}$ correspond to the annihilation operators for incoming and losses modes of the cavity which have zero mean value, except for the incoming pump ($\left<\hat{a}_p^{in}(t)\right>=\!\!\sqrt{P_{in}/(\hbar\Omega_p)}$), and their fluctuations are delta-correlated $\left<\delta\hat{a}_j^{loss/in \dagger}(t)\delta\hat{a}_k^{loss/in}(t^\prime)\right>\!=\!\delta_{j,k}\delta(t-t^\prime)$. Finally, we have that the incoming, intra-cavity and output fields are related by the usual beam splitter relation $\hat{a}^{out}=-\hat{a}^{in}+\sqrt{2\gamma}\hat{a}$.

We proceed by linearizing the quantum fluctuations, $\hat{a}=\alpha+\delta\hat{a}$. The mean values $\left\langle\hat{a}_j\right\rangle=\alpha_j$, satisfy the classical equations of motion which correspond to the substitution $\hat{a}\rightarrow\alpha$ in (\ref{eqn:chi3EqnMovModes}). Following \cite{Matsko2005} we solve these classical equations in the steady state by setting $\alpha_j=A_je^{i\theta_j}$ and accounting for the real and imaginary parts. Two coupled algebraic equations for the amplitudes $A=A_s=A_i$ and $A_p$ are obtained in terms of the experimentally adjustable parameters, the normalized input pump power $F^2=\frac{2\gamma_p \eta}{\hbar \Omega_p \Gamma^3} P_{in}$ and its detuning $\Delta_p$.
\begin{gather}
A_p^4=1+\left(\Delta_p-\frac{D_3}{2}-2A_p^2-3A^2\right)^2\nonumber\\
F^2\!=\!A_p^2\left\{\!\left(\!1+2\frac{A^2}{A_p^2}\!\right)^2\!\!+\!\!\left[\Delta_p-A_p^2-\frac{2A^2}{A_p^2}\left(\Delta-3A_p^2\right)\right]^2\right\}.\label{eqn:AnalyticalSol3M}
\end{gather}

\begin{figure}[htbp]
	\begin{center}
		\includegraphics[width=.49\linewidth]{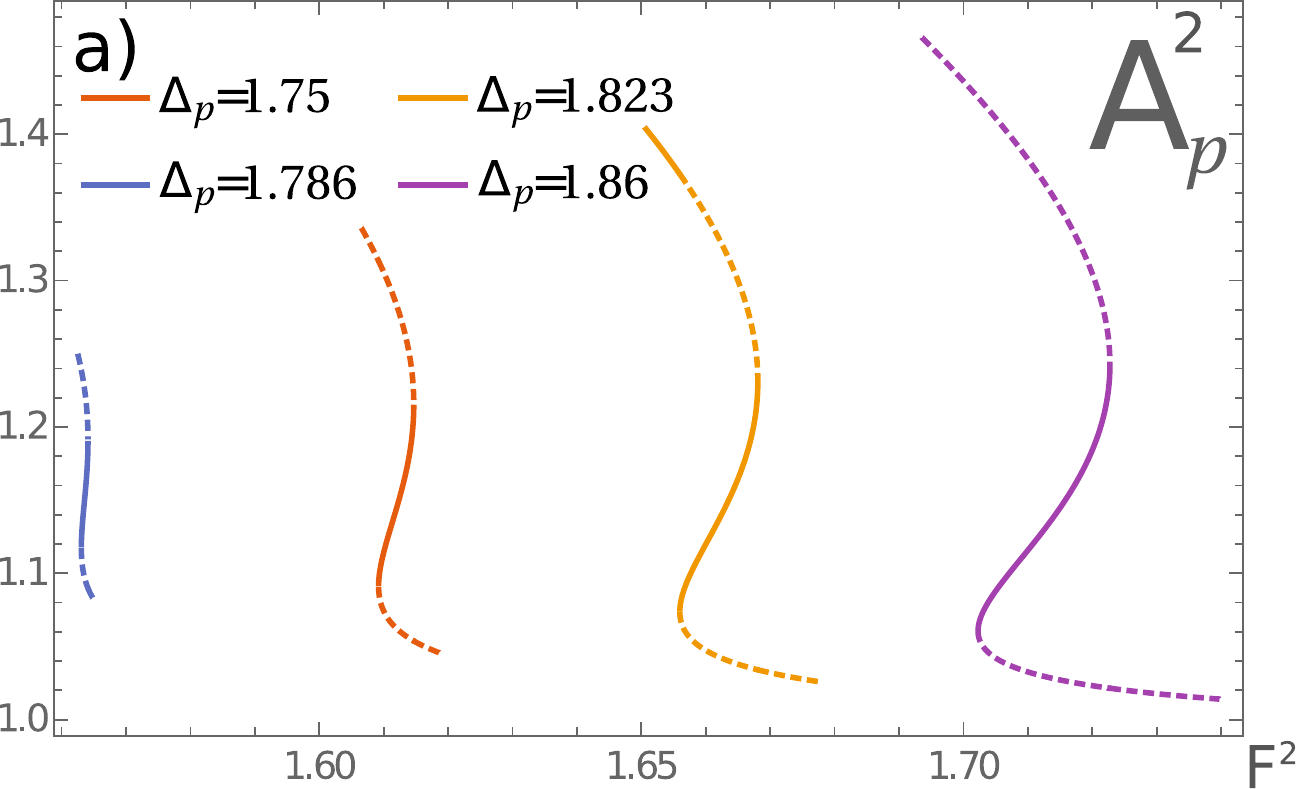}
		\includegraphics[width=.49\linewidth]{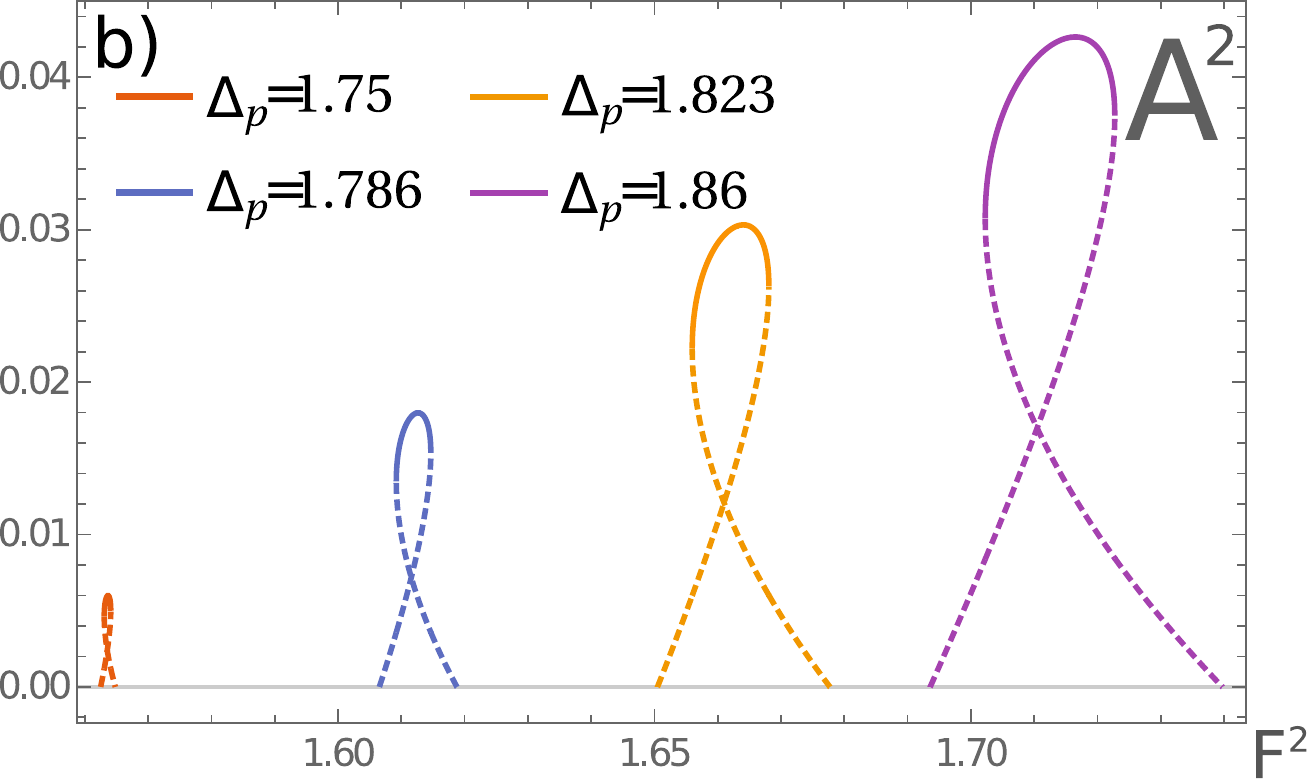}
		\includegraphics[width=.49\linewidth]{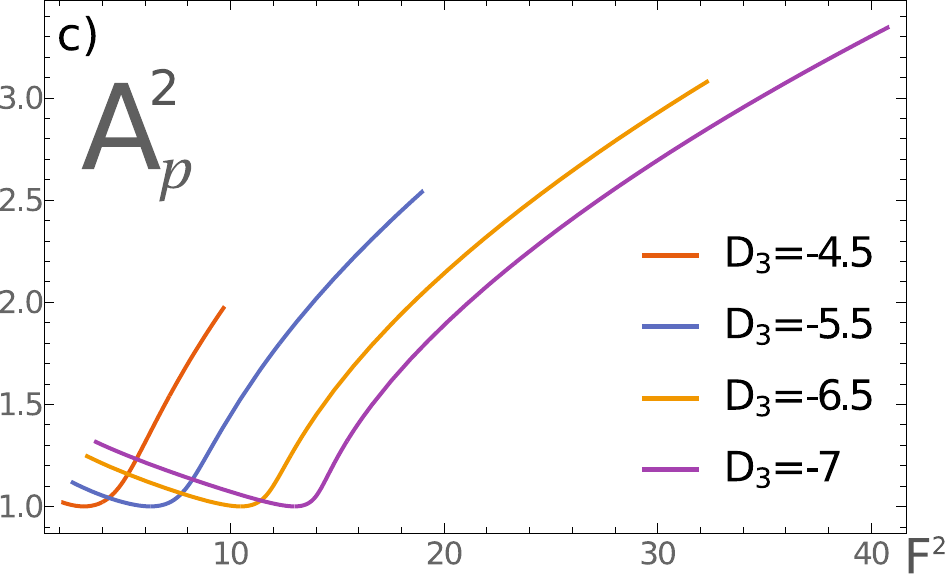}
		\includegraphics[width=.49\linewidth]{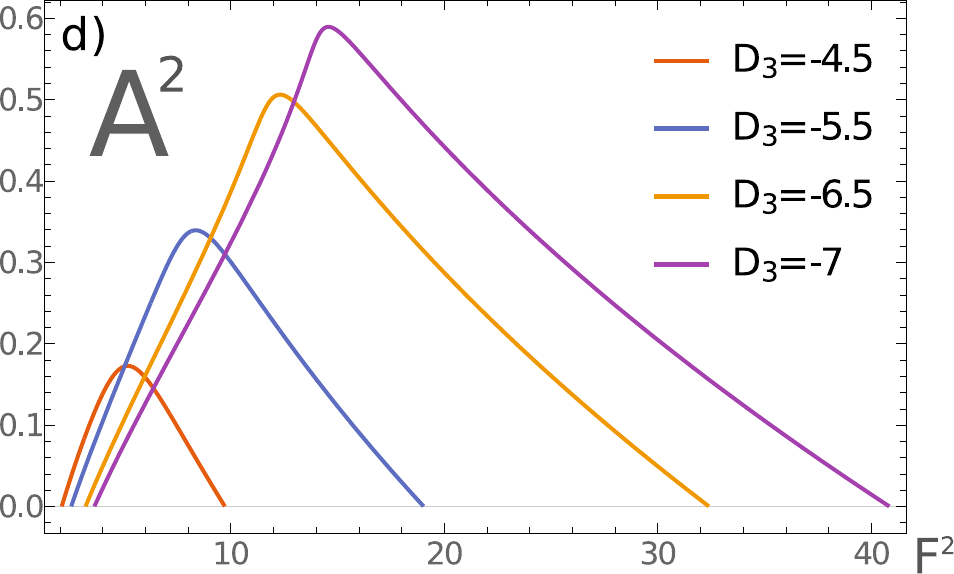}
		\caption{Intracavity pump (left) and signal-idler (right) amplitudes for several values of pump detuning $\Delta_p$ at perfect phase matching $D_3=0$ (top) and on resonance with the pump, $\Delta_p=0$ for several values of phase matching $D_3$ (bottom). The dashed lines represent unstable solutions.}
		\label{fig:DepltionMeanChi3}
	\end{center}
\end{figure}

We work with normalized quantities replacing $A\rightarrow\sqrt{\frac{\Gamma}{\eta }}A;\:D_3\rightarrow\frac{D_3}{\Gamma};\:\Delta_p\rightarrow\frac{\Delta_p}{\Gamma}$. The numerical solutions for (\ref{eqn:AnalyticalSol3M}), in the case of perfect phase matching ($D_3=0$) for different values of pump detuning $\Delta_p$ are shown in Fig.~(\ref{fig:DepltionMeanChi3} a, b), while the case of perfect pump resonance ($\Delta_p=0$) for several values of $D_3$ are plotted in Fig.~(\ref{fig:DepltionMeanChi3} c, d). On the left (a, c) we have the normalized power of the intracavity pump $A_p^2$ and on the right (b, d) the signal and idler normalized power $A^2$.

Consequences of phase modulation are evident in Fig. \ref{fig:DepltionMeanChi3}. For instance, the intracavity pump field does not have a constant value above threshold, in contrast to the $\chi^{(2)}$ scenario \cite{Debuisschert1993,Fabre1997} depending instead on the incoming power. Exactly on resonance and for perfect phase matching (Fig \ref{fig:DepltionMeanChi3} a and b), there is no oscillation. This is because when we increase the pump power in order to trigger the oscillation, the phase modulation shifts the mode frequencies out of resonance. It is necessary to pump the resonator with nonzero detuning (even larger than the cavity linewidth) to compensate this effect. In the case of perfect resonance this compensation is done by the phase mismatch (Fig \ref{fig:DepltionMeanChi3} c and d). For the same reason, our model predicts that for higher values of $F^2$ the oscillation ceases. In most experiments, however, higher pump powers excite other cavity modes and to the generation of frequency combs. This feature is not predicted in the model given that only three modes are taken into account.

Noise properties are calculated by writing (\ref{eqn:chi3EqnMovModes}) up to first order in quantum fluctuations $\delta\hat{a}_i$. Defining a vector of fluctuations as 
\begin{eqnarray*}
	\delta\mathbf{{\hat{{A}}}}=\{\delta \hat{a}_p e^{-i\theta_p},\delta \hat{a}_p^{\dagger}e^{i\theta_p},\delta \hat{a}_se^{-i\theta_s},\delta \hat{a}_s^\dagger e^{-i\theta_s},\delta \hat{a}_ie^{-i\theta_i},\delta \hat{a}_i^\dagger e^{-i\theta_i}\}^T,
\end{eqnarray*} 
where $\theta_i$ is the mean field phase $\alpha_i=A_ie^{i\theta_i}$, the evolution of these fluctuations is given by
\begin{gather}
\frac{d \delta\hat{\mathbf{A}}}{d\tau}=-\mathbb{M}_a\cdot\delta\hat{\mathbf{A}}+\frac{\mathbb{T}_a^{in}}{\Gamma}\cdot\delta\hat{\mathbf{A}}^{in}+\frac{\mathbb{T}_a^{loss}}{\Gamma}\cdot\delta\hat{\mathbf{A}}^{loss}\label{eqn:MotiondA}\\
\mathbb{T}_a^{in}=\text{Diag}\{\sqrt{2\gamma_p}e^{-i\theta_p},\sqrt{2\gamma_p}e^{i\theta_p},\sqrt{2\gamma},\sqrt{2\gamma},\sqrt{2\gamma},\sqrt{2\gamma}\},\nonumber 
\end{gather}
with $\mathbb{T}_a^{loss}$ defined in the same way as $\mathbb{T}_a^{in}$ but changing $\gamma$ by $\mu$ and where $\tau=t\Gamma$ is the normalized time. The matrix $\mathbb{M}_a$ is derived from the linearization process and its elements are functions of the field's mean values and the detunings.

As a first step, we treat the pump classically ($\delta \hat{a}_p\rightarrow0$) and study only the correlations between the signal and idler fields. The observables of interest are the quadrature operators of each field $\hat{x}=(\hat{a}+\hat{a}^\dagger)/\sqrt{2}$ and $\hat{y}=-i(\hat{a}-\hat{a}^\dagger)/\sqrt{2}$ referred to as intensity and phase quadratures, which satisfy the commutation relation $\left[\hat{x},\hat{y}\right]=i$.

\begin{figure}[h]
	\begin{center}
		\includegraphics[width=.47\linewidth]{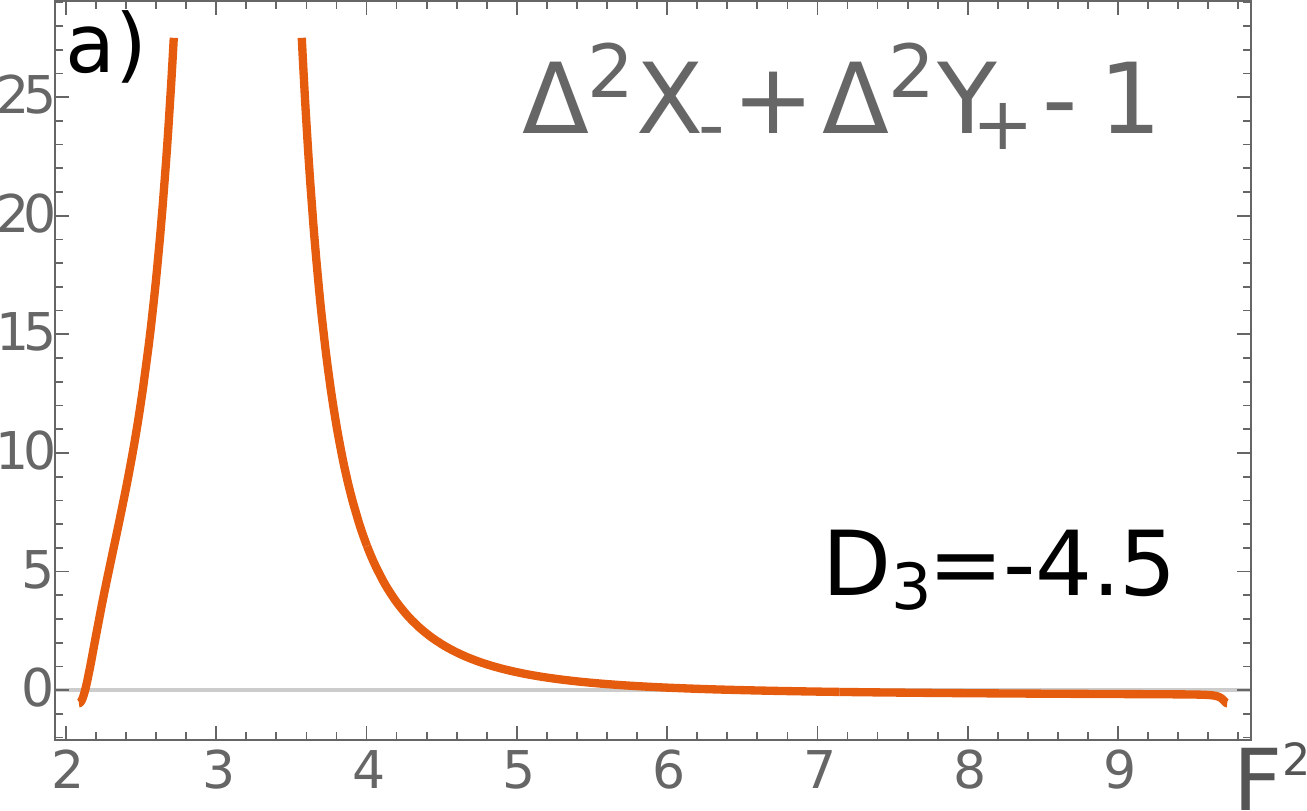}
		\includegraphics[width=.47\linewidth]{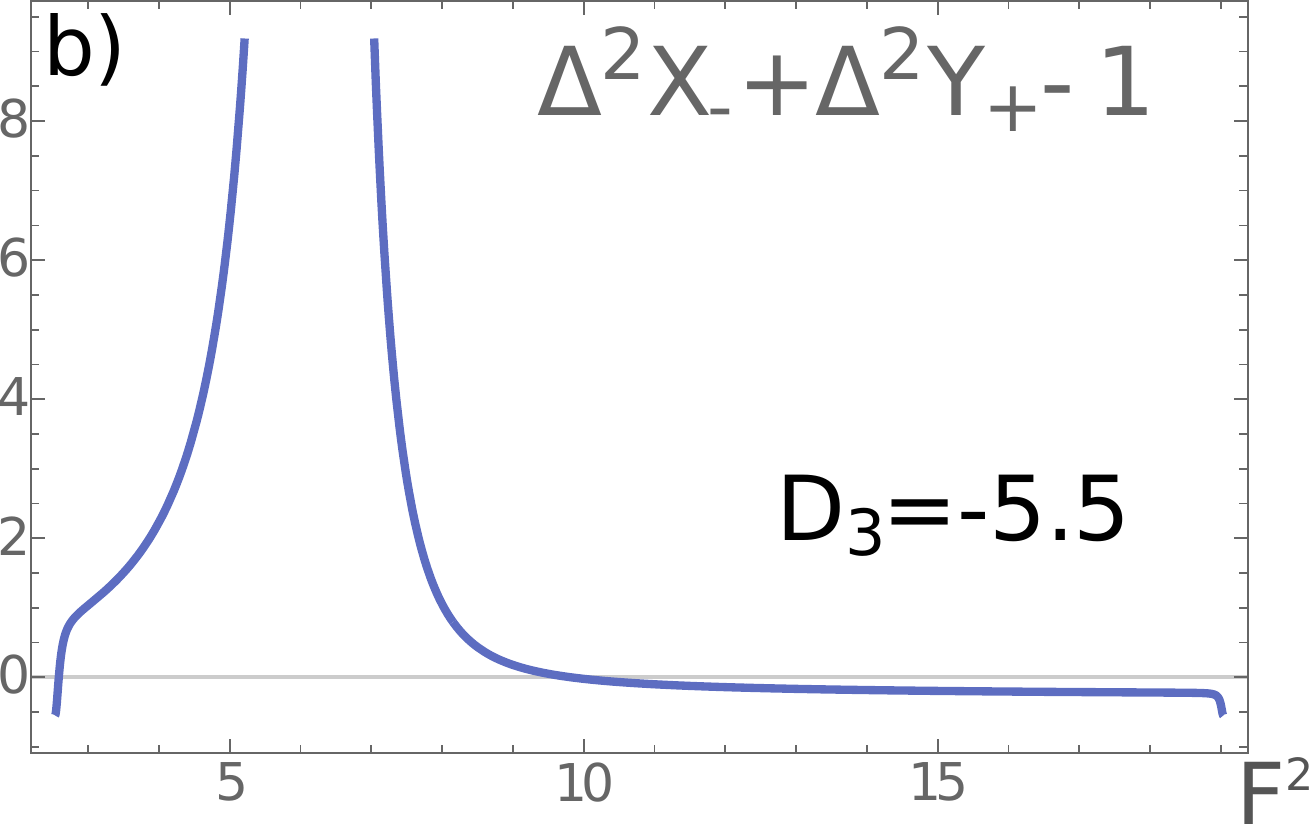}
		\includegraphics[width=.47\linewidth]{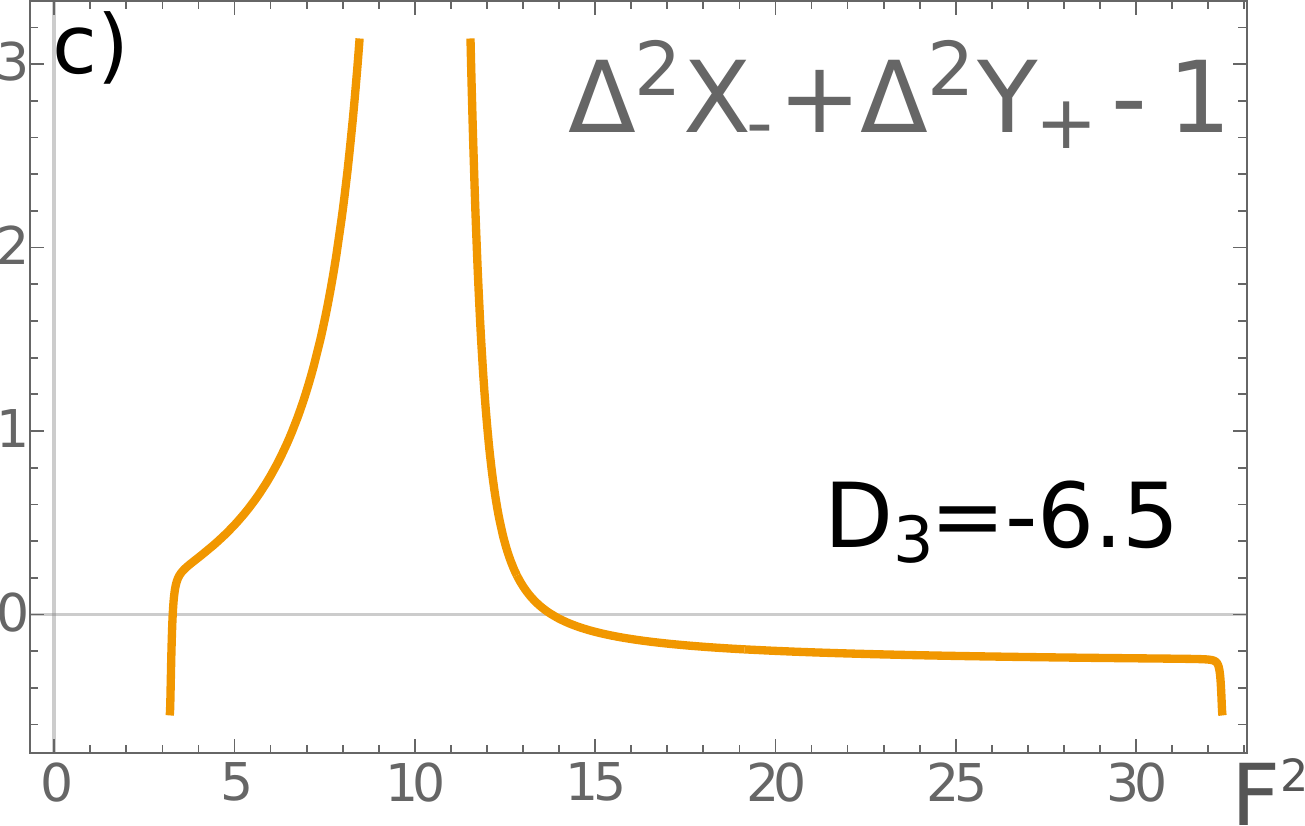}
		\includegraphics[width=.47\linewidth]{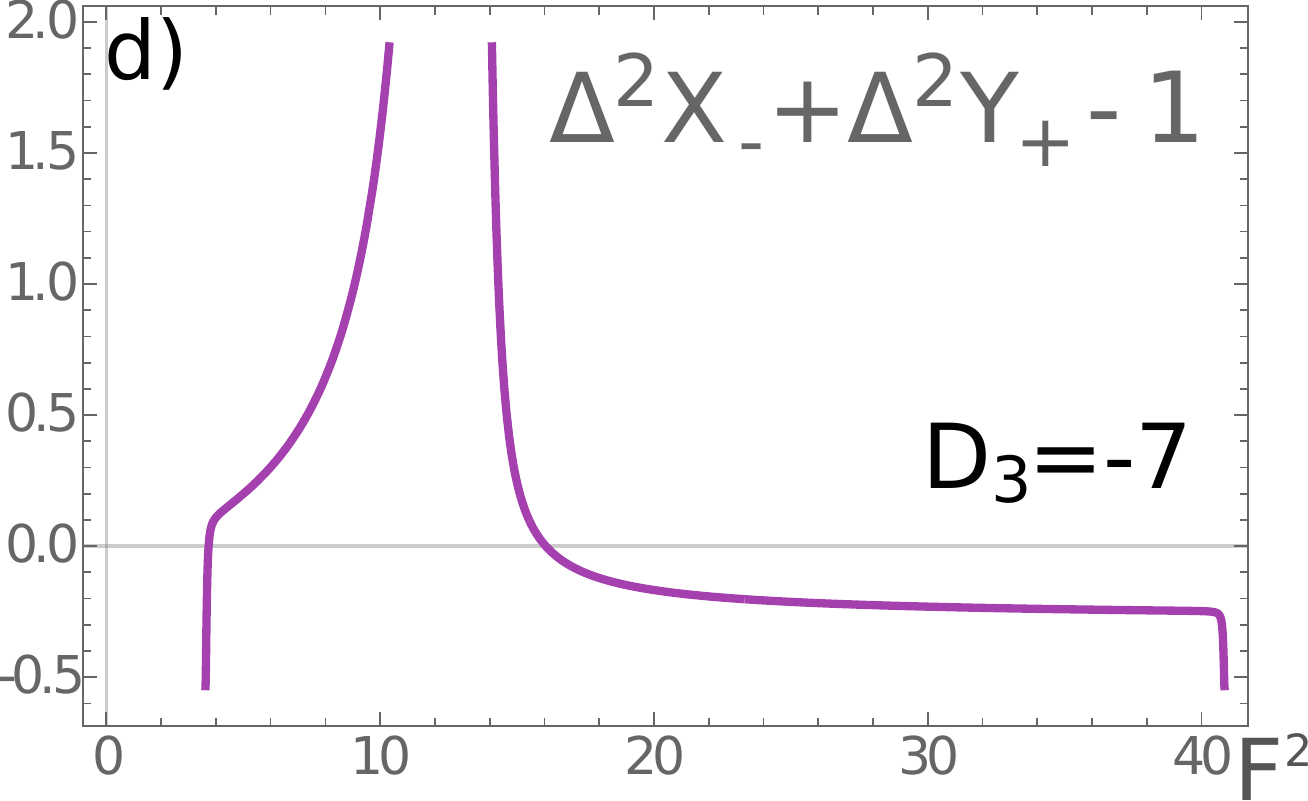}
		\caption{\textit{L.h.s} of Duan inequality (\ref{eqn:DuanIqn}) for $\Delta_p=0$ and different values of $D_3$. Negative values signal entanglement in the system.}
		\label{fig:DuanChi3}
	\end{center}
\end{figure}

Equations (\ref{eqn:MotiondA}) decouple if we write them in the sum/subtraction basis. Thus, the vector $\delta\hat{\mathbf{X}}_{pm}=\{\delta\hat{y}_+,\delta\hat{x}_+,\delta\hat{y}_-,\delta\hat{x}_-\}^T$, with $\hat{x}_\pm=(\hat{x}_s\pm\hat{x}_i)/\sqrt{2}$ and $\hat{y}_\pm=(\hat{y}_s\pm\hat{y}_i)/\sqrt{2}$, fulfills the equation
\begin{eqnarray}
\frac{\delta\hat{\mathbf{X}}_{pm}}{d\tau}=\mathbb{M}_{pm}\cdot\delta\hat{\mathbf{X}}_{pm}+\frac{\mathbb{T}_{pm}^{in}}{\Gamma}\cdot\delta\hat{\mathbf{X}}_{pm}^{in}+\frac{\mathbb{T}_{pm}^{loss}}{\Gamma}\cdot\delta\hat{\mathbf{X}}_{pm}^{loss}\nonumber\\
	\mathbb{M}_{pm}=\text{diag}\{\mathbb{M}_{+},\mathbb{M}_-\},
\label{eqn:FluctiationEvoPM}
\end{eqnarray}
were $\mathbb{M}_{+}$ and $\mathbb{M}_-$ are $2\times2$ matrices.
Next, we calculate the noise spectral density matrix $S(\omega)$ as $S(\omega)\delta(\omega+\omega^\prime)=\left\langle\delta\hat{X}_{pm}(\omega)\delta\hat{X}_{pm}(\omega^\prime)^T\right\rangle$ from which we can extract all the noise and entanglement information. In particular, we can evaluate the Duan inequality \cite{Duan2000} to test entanglement between signal and idler fields
\begin{gather}
\Delta^2\hat{x}_- + \Delta^2\hat{y}_+  -1\geq 0.\label{eqn:DuanIqn}
\end{gather}
The inequality (\ref{eqn:DuanIqn}) is presented in Fig.~\ref{fig:DuanChi3} for different values of $D_3$ as a function of $F^2$. Negative values of this quantity (i.e. violation of Duan inequality) imply entanglement between signal and idler. Entanglement detected using (\ref{eqn:DuanIqn}) is highly limited to a small region for high values of $F^2$. Here, typical experimental parameters were used, $\omega/\Gamma=1.5\times10^{-2}$ and $\gamma/\Gamma=0.55$ which correspond to an analysis frequency of 3 MHz, a bandwidth of 200 MHz, intrinsic and loaded $Q$s of $10^6$ and 450.000, respectively, for $\lambda_p=1561nm$.

In contrast to PDC-based OPOs, where the Duan inequality has been proven as an excellent entanglement witness, we observe the effects of phase modulation here. As pointed out by Ferrini {\em et al.}~\cite{Ferrini2014}, the phase modulation distorts the noise ellipse and thus the strongest correlations are not between the difference of intensities $\hat{x}_-$ and the sum of the \textit{phases} $\hat{y}_+$. Stronger correlations exist among other quadratures to be determined.

We proceed by calculating the Schmidt modes, which provide the best attainable squeezing for the system~\cite{Ferrini2014}. The quadratures of theses Schmidt modes (denoted here with the superscript $rot$) are given by a \textit{rotation} of the quadratures in the following way
\begin{eqnarray}
\left(\begin{array}{c}
y^{rot}_\pm\\
x^{rot}_\pm
\end{array}\right)
=\left(
\begin{array}{cc}
\cos (\theta_\pm ) & \sin (\theta_\pm ) \\
-\sin (\theta_\pm ) & \cos (\theta_\pm ) \\
\end{array}
\right)\left(\begin{array}{c}
y_\pm\\
x_\pm
\end{array}\right)\label{eqn:RotationPlusMinus}
\end{eqnarray} such that the noise spectral density matrix is diagonal. 
The Duan inequality for testing signal-idler entanglement using the rotated quadrature can be rewritten in terms of the Schmidt mode quadratures as~\cite{Giovannetti2003}
\begin{gather}
\Delta^2\hat{x}^{rot}_-  + \Delta^2\hat{y}_+^{rot} -\vert C \vert\geq 0\label{eqn:SchmidtDuanIqn}\quad
\text{with}\quad C=\cos(\theta_+-\theta_-).
\end{gather}

\begin{figure}[htb]
	\begin{center}
		\includegraphics[width=.8\linewidth]{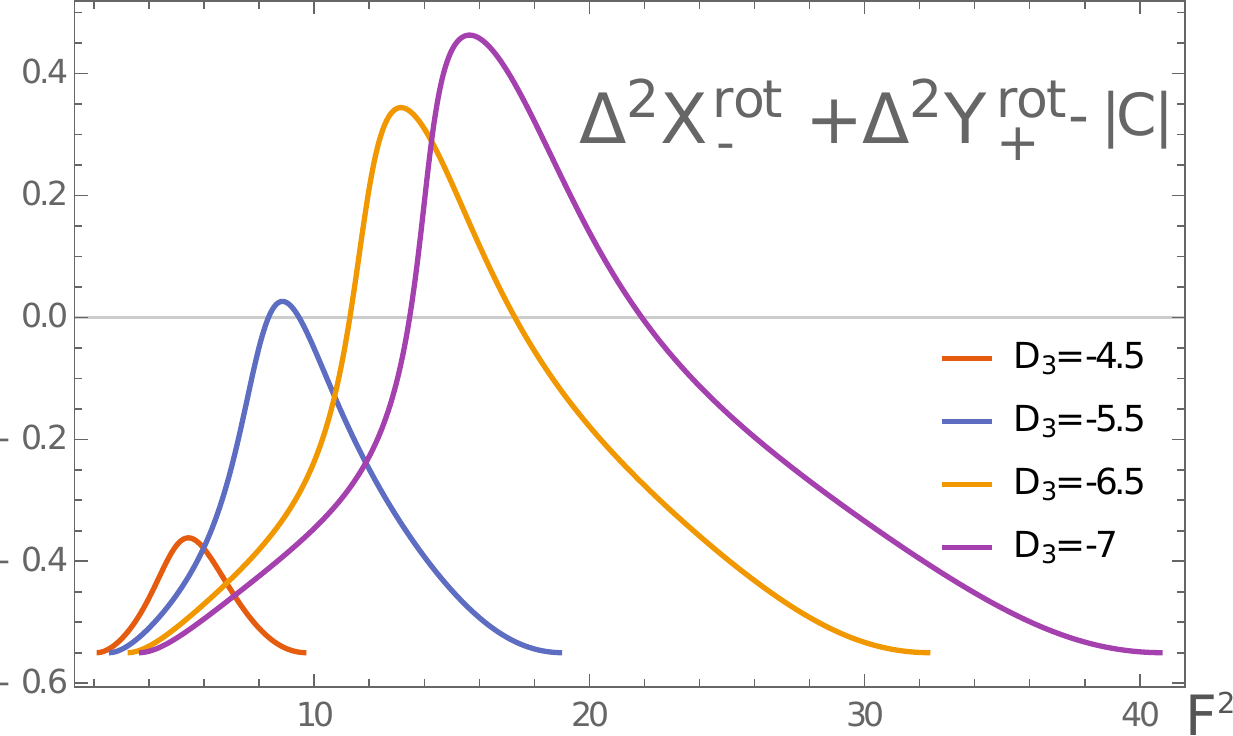}
		\caption{Detection of entanglement between signal and idler fields using the Duan inequality written in terms of Schmidt modes (\ref{eqn:SchmidtDuanIqn}), for $\Delta_p=0$ and several values of phase matching $D_3$. Values below $0$ indicate entanglement.}
		\label{fig:DuanRotChi3}
	\end{center}
\end{figure}

\begin{figure}[htbp]
	\begin{center}
		\includegraphics[width=.5\linewidth]{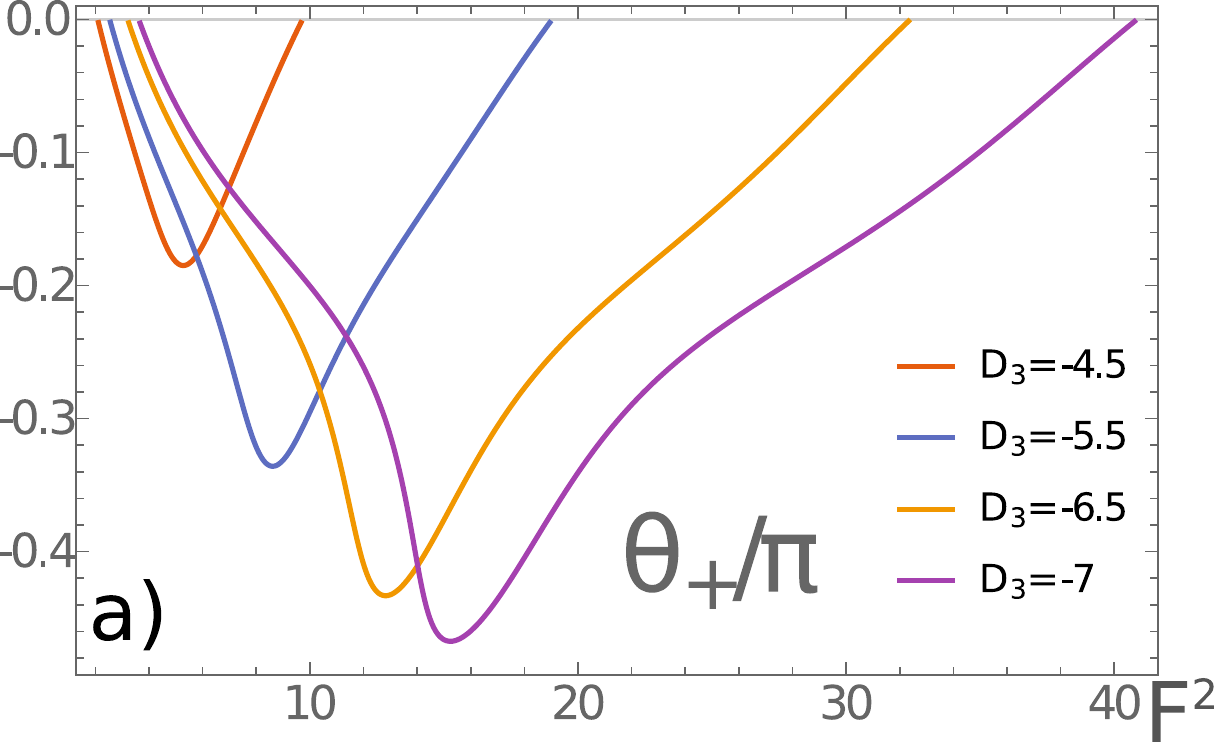}
		\includegraphics[width=.47\linewidth]{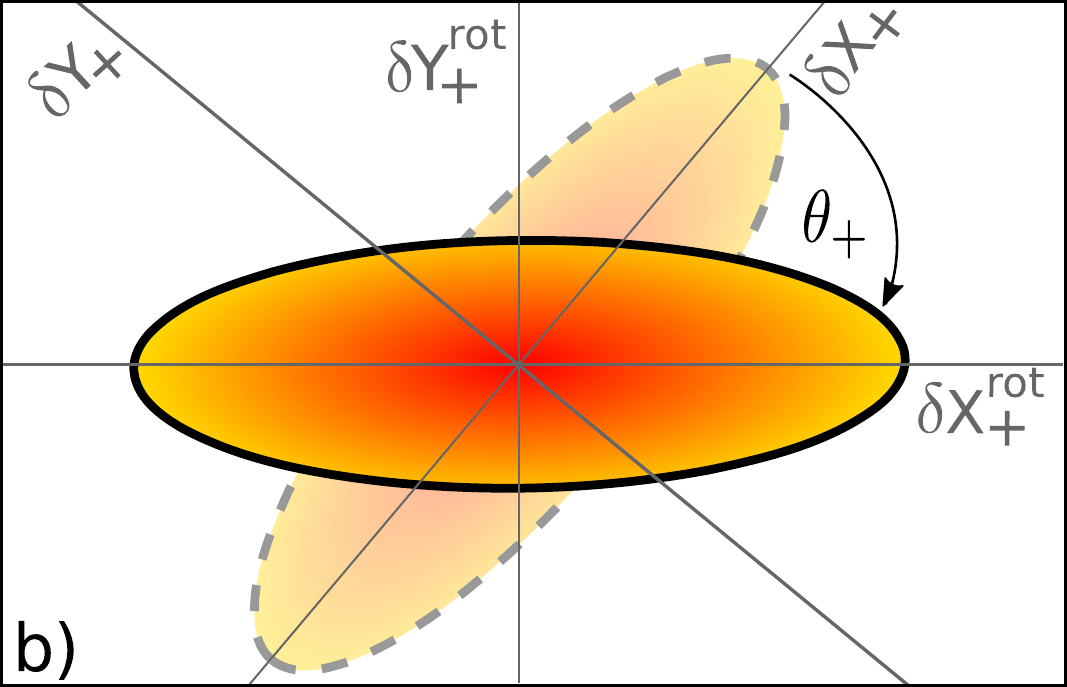}
		\caption{ a) Rotation angle for the sum subspace as a function of the incoming power with $\Delta_p=0$ for several values of phase matching $D_3$. b) Representation of the effect of the quadrature rotation. The noise ellipse axis is aligned with the $\{\delta X_+,\delta Y_+\}$ axis.}
		\label{fig:ThetaPlus}
	\end{center}
\end{figure}

The modified Duan inequality is displayed in Fig.~\ref{fig:DuanRotChi3}, indicating that the parameter region where entanglement is detected increases appreciably and there is only a small region where no entanglement is detected.
 We note that the rotation for the sum and subtraction quadratures are independent. The subtraction rotation is negligible, $|\theta_-/\pi|<10^{-5}$, but the rotation of the quadrature sum is significant (see figure \ref{fig:ThetaPlus}), providing a better quadrature to detect entanglement. The regions where the left hand side of (\ref{eqn:SchmidtDuanIqn}) is bigger coincide with the region of larger rotation of the sum quadrature. In particular, the inequality is not violated for $\theta_+\approx\pi/2$ which implies $\hat{y}^{rot}_+\approx\hat{x}_+$ i.e. only intensity information is taken into account and, therefore, the inequality must be fulfilled.

Next we analyze the three mode system by taking into account the quantum fluctuations of the pump field. Defining the vector $\delta\hat{\mathbf{X}}=\{\delta\hat{y}_p,\delta\hat{x}_p,\delta\hat{y}_+,\delta\hat{x}_+,\delta\hat{y}_-,\delta\hat{x}_-\}$, its linearized equation of motion is given by
\begin{eqnarray*}
	\frac{\delta\hat{\mathbf{X}}}{d\tau}=\mathbb{M}\cdot\delta\hat{\mathbf{X}}+\frac{\mathbb{T}^{in}}{\Gamma}\cdot\delta\hat{\mathbf{X}}^{in}+\frac{\mathbb{T}^{loss}}{\Gamma}\cdot\delta\hat{\mathbf{X}}^{loss} \;.
\end{eqnarray*}
In this situation the pump fluctuations couple only to the sum subspace, while the subtraction remains the same as in the previous case. The $\mathbb{M}$ matrix is block diagonal $\mathbb{M}=$diag$\{\mathbb{M}_{^p_+},\mathbb{M}_{-}\}$, where $\mathbb{M}_{^p_+}$ is the matrix coupling the sum and pump modes and  $\mathbb{M}_{-}$ is the same as in (\ref{eqn:FluctiationEvoPM}) (explicit expressions of these matrices can be found in \cite{Gonzalez2017}). In order to analyze the entanglement between three modes, we use the van Loock-Furusawa criteria \cite{vanLoock2003}. We resort again to the Schmidt modes to find the linear combination which is best to detect entanglement. Schmidt quadratures $\hat{\boldsymbol{\xi}}=\{\hat{\xi}_1,\hat{\xi}_2,\hat{\xi}_3,\hat{\xi}_4\}^T$ are determined by
\begin{gather}
\hat{\boldsymbol{\xi}}=\mathbb{U}\cdot \hat{\mathbf{X}}_{_p^+} \label{eqn:RotationPumpPlus}
\end{gather}
with $\hat{\mathbf{X}}_{_p^+}=\{\hat{x}_p,\hat{y}_p,\hat{x}_+,\hat{y}_+\}^T$ and where $\mathbb{U}$ is the Schmidt transformation (unitary) relating the Schmidt modes with the original modes and diagonalizes the noise spectral density matrix. As long as the subtraction fluctuations are not affected by the pump noise, their corresponding Schmidt quadratures are the same as in (\ref{eqn:RotationPlusMinus}). In addition to being of higher dimension, the matrix $\mathbb{U}$ can also be complex and can not be seen as a geometrical rotation in a 4-dimensional space as in the previous case.
In terms of the Schmidt quadratures with lower noise, the van Loock-Furusawa inequalities take the form
\begin{eqnarray}
	\Delta^2 \xi_i+\Delta^2 \xi_j-\left|S_{I}^{i,j}\right|\geq 0\quad\text{and}\quad
	\Delta^2 x_-^{rot}+\Delta^2 \xi_j-\left|P_{I}^{j}\right|\geq 0 \;, \label{eqn:VanlockFIneqn} 
\end{eqnarray}
where $S_{I}^{i,j}$ and $P_{I}^{j}$ are separability constants which depend on the quadratures $\xi_i$ ($i=3,4$, the quadratures with less noise) and $x_-^{rot}$ used, the bipartition $I$ in which the inseparability is being tested, and are functions of the matrices in (\ref{eqn:RotationPlusMinus}) and (\ref{eqn:RotationPumpPlus}). Since the system is symmetric on the interchange of the signal and idler field, the separability constants are the same for the bipartitions $I_1=\{s|i,p\}$  and $I_2=\{i|s,p\}$.
\begin{figure}[h]
	\begin{center}
		\includegraphics[width=.47\linewidth]{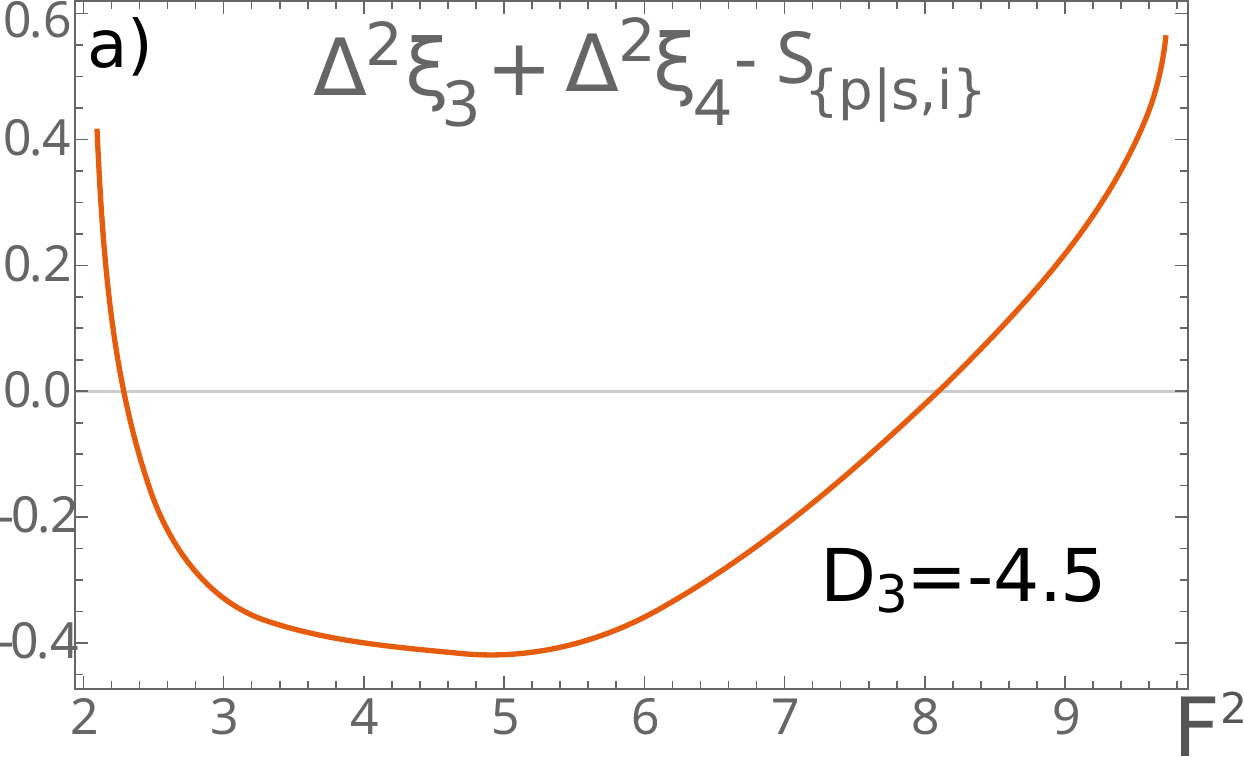}
		\includegraphics[width=.47\linewidth]{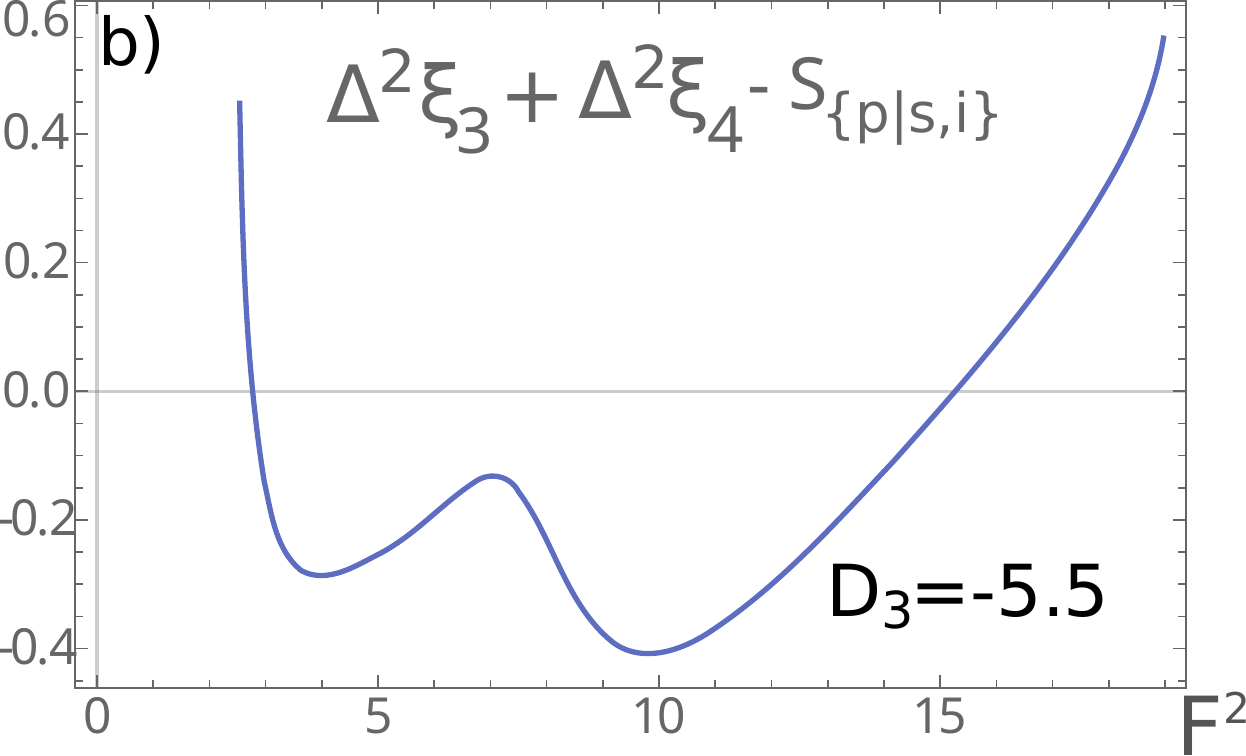}
		\includegraphics[width=.47\linewidth]{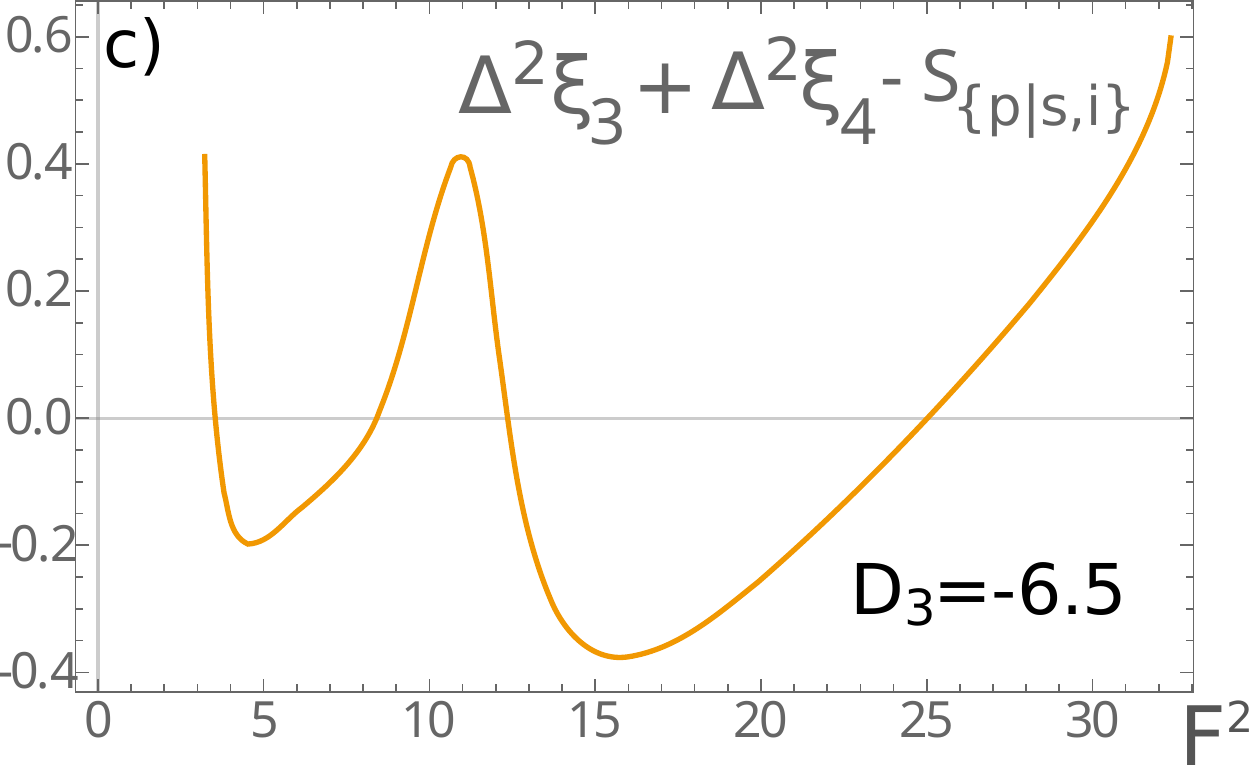}
		\includegraphics[width=.47\linewidth]{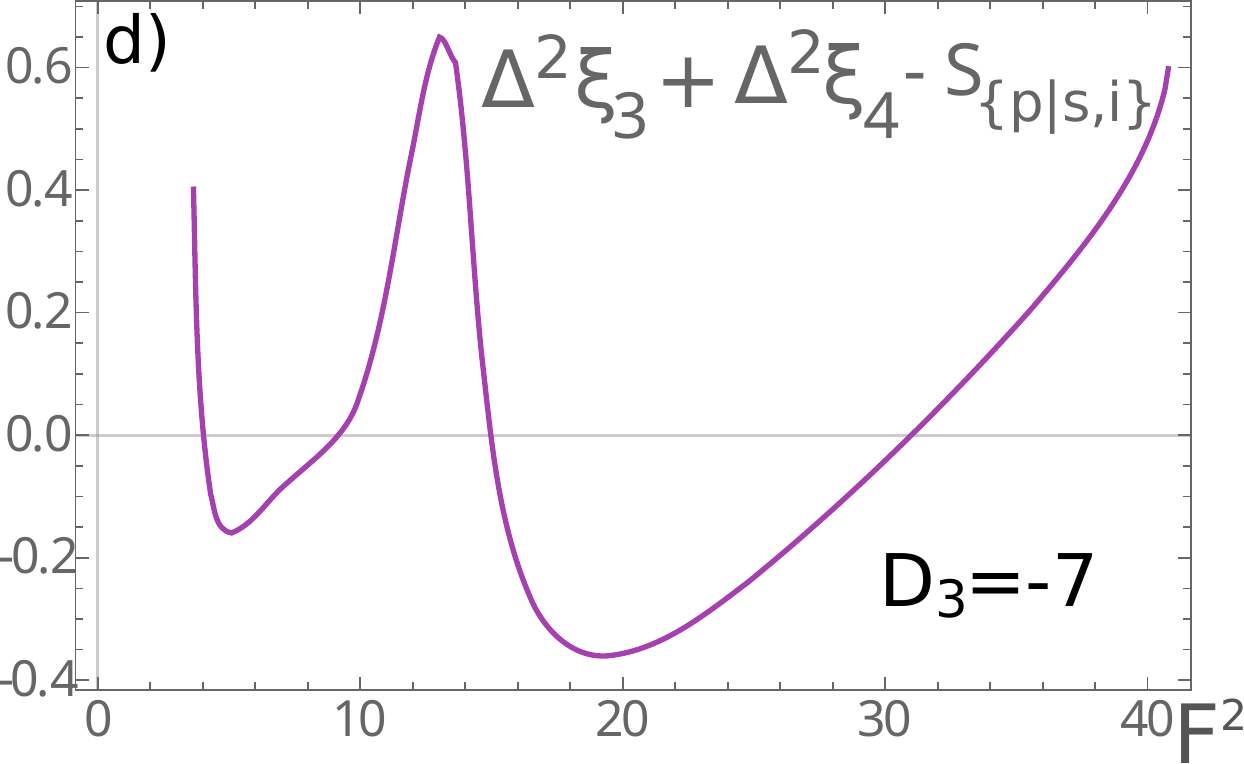}
		\caption{Inseparability test for partition $I=\{p|s,i\}$}
		\label{fig:3MInseparability1}
	\end{center}
\end{figure}
\begin{figure}[h]
	\begin{center}
		\includegraphics[width=.47\linewidth]{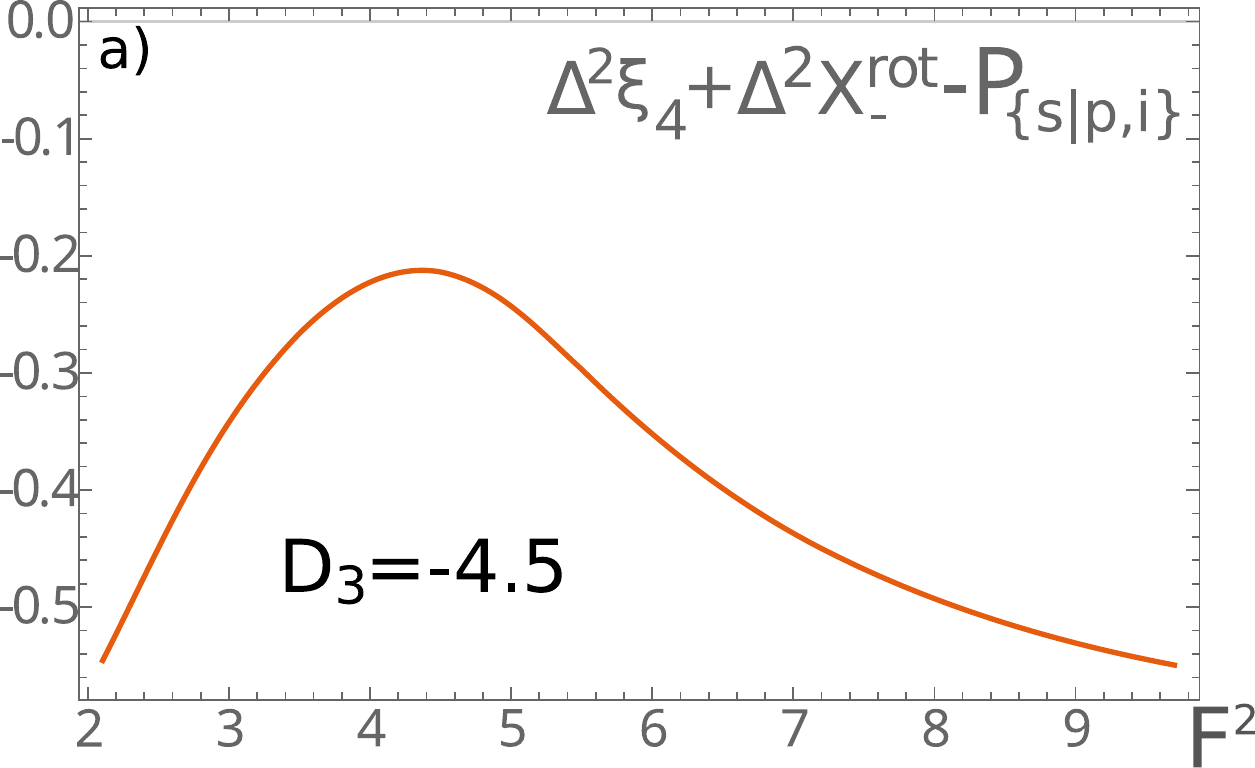}
		\includegraphics[width=.47\linewidth]{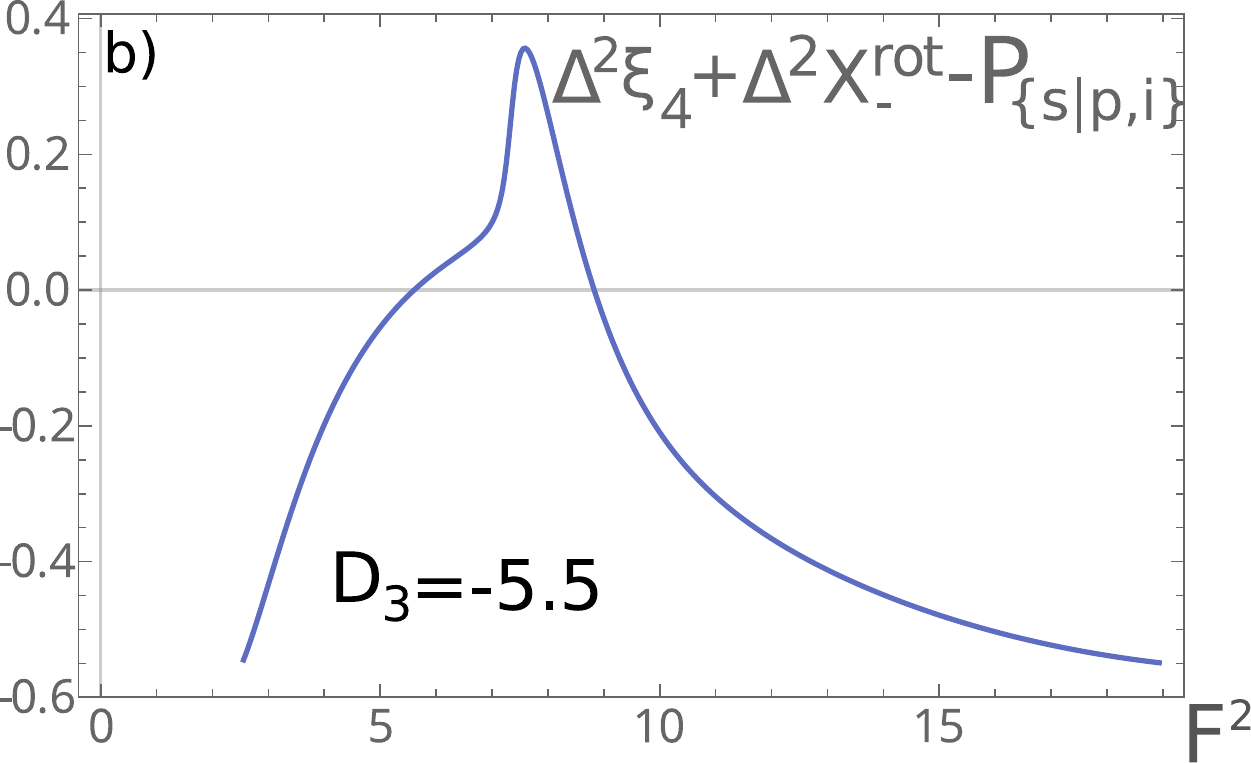}
		\includegraphics[width=.47\linewidth]{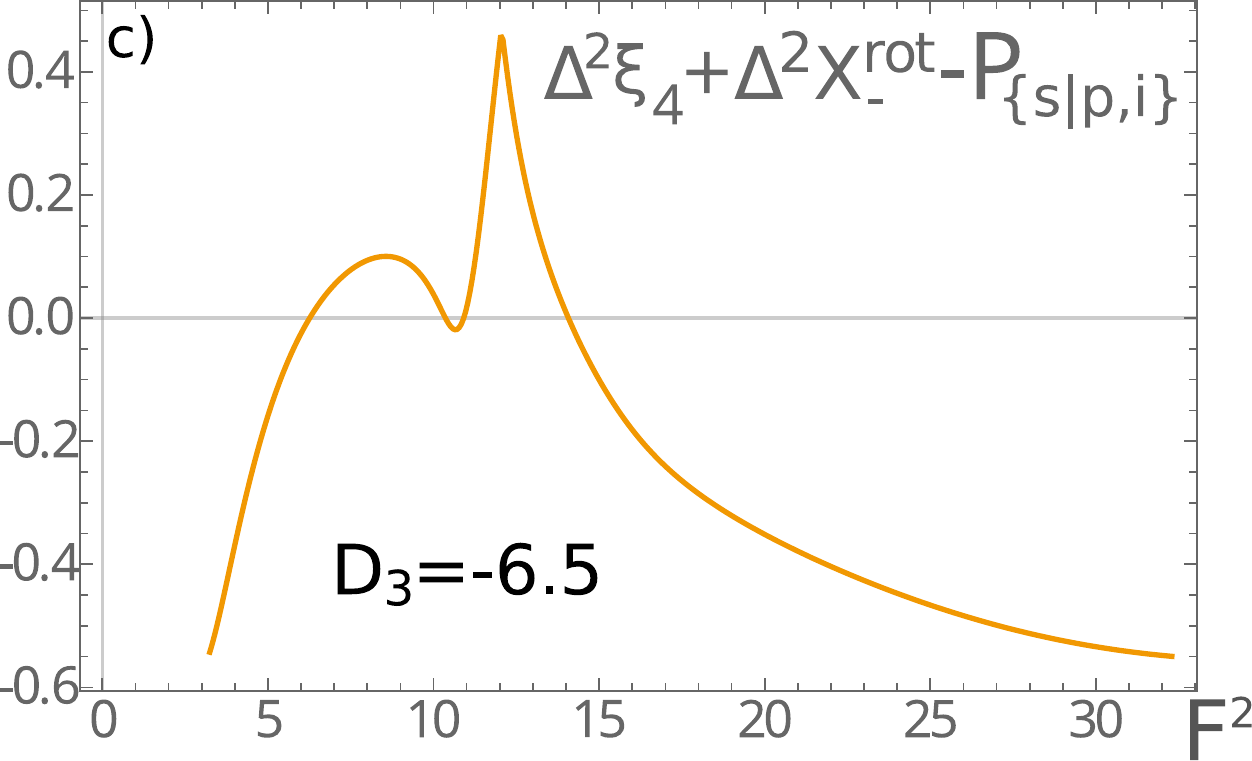}
		\includegraphics[width=.47\linewidth]{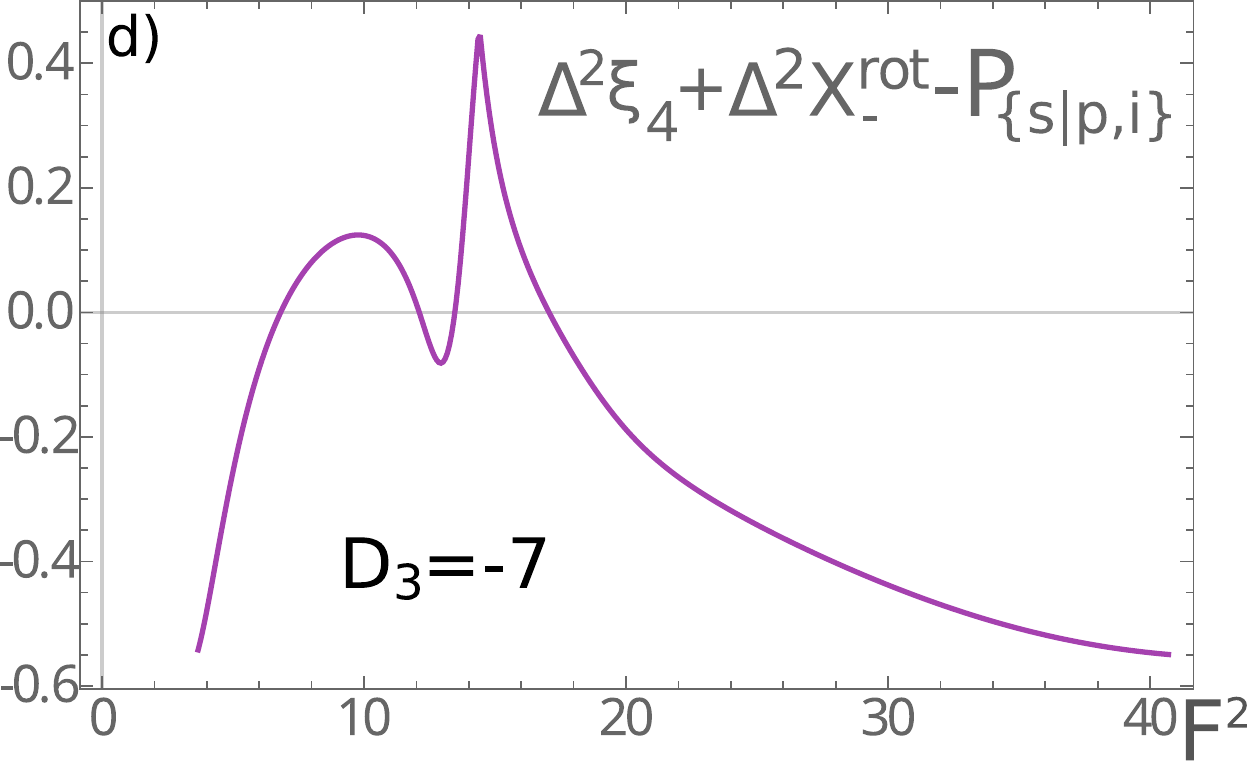}
		\caption{Inseparability test for partitions $I=\{s|i,p\}=\{i|s,p\}$}
		\label{fig:3MInseparability2}
	\end{center}
\end{figure}

Fig. \ref{fig:3MInseparability1} and fig. \ref{fig:3MInseparability2} show the \textit{l.h.s.} of the two inequalities of the form (\ref{eqn:VanlockFIneqn}) that are violated for the Schmidt quadratures. The violation of the inequality (negative values) plotted in \ref{fig:3MInseparability1} demonstrates entanglement between the system formed by the pump mode on one side and the signal and idler modes on the other side.  Given the symmetry of the system, the violation of the inequality plotted in \ref{fig:3MInseparability2} demonstrates inseparability of the signal from the pump-idler modes as well as inseparability of the idler from the pump-signal modes.

In summary, we have provided the first complete theoretical analysis of three-mode correlations in a $\chi^{(3)}$ OPO operating above-threshold, taking into account phase modulation effects as well as mode dispersion inherent to on-chip realizations. We predict tripartite pump-signal-idler entanglement, as observed in the $\chi^{(2)}$ OPO~\cite{Coelho2009}. On-chip generation of continuous-variable multipartite entangled states shall be especially useful for ultrashort distance quantum communications on the scale of future quantum processors.

\textbf{Funding}.This work was supported the São Paulo Research Foundation (FAPESP), CNPq and CAPES (Brazilian agencies).

\textbf{Acknowledgment.} The authors thank professor M. Martinelli for use useful discussions.

\bibliography{biblio}


\end{document}